\documentclass[12pt, prd, showpacs]{revtex4}
\usepackage{amssymb}
\usepackage{amsmath}

\input{tcilatex}

\begin{document}

\title{Kinematics of ultra-high energy particle collisions near black holes
in the magnetic field }
\author{O. B. Zaslavskii}
\email{zaslav@ukr.net }
\affiliation{Department of Physics and Technology, Kharkov V.N. Karazin National
University, 4 Svoboda Square, Kharkov 61022, Ukraine}
\affiliation{Institute of Mathematics and Mechanics, Kazan Federal University, 18
Kremlyovskaya St., Kazan 420008, Russia}

\begin{abstract}
There are different versions of collisions of two particles near black holes
with unbound energy $E_{c.m.}$ in the centre of mass frame. The so-called
BSW effect arises when a slow fine-tuned "critical" particle hits a rapid
"usual" one. We discuss a scenario of collision in the strong magnetic field
for which explanation tuns out to be different. Both particles are rapid but
the nonzero angle between their velocities (which are both close to $c$, the
speed of light) results in a relative velocity close to $c$ and, hence, big $%
E_{c.m.}$.
\end{abstract}

\maketitle

\section{Introduction}

Several years ago, an interesting observation was made that the collision
between two particles near the extremal Kerr metrics can lead to the unbound
growth of the energy in their centre of mass $E_{c.m.}$\cite{ban} although
individual Killing energies $E_{1}$ and $E_{2}$ are finite. This effect is
now called the Ba\~{n}ados - Silk - West (BSW) effect. This result came as a
surprise but later it was shown that the BSW effect reveals itself for quite
generic black holes and has a simple kinematic explanation \cite{k}. The BSW
effect occurs if one particle is so-called critical (has fine-tuned energy
and angular momentum or electric charge) while the second particle is
"usual" (not fine-tuned). Then, in the stationary or static reference frame,
a usual particle moves near the horizon with a speed approaching that of
light, whereas the velocity of the critical one is separated from it. Then,
a very rapid particle hits a slow target that results in \ a large Lorentz
factor of relative motion, hence large $E_{c.m.}$.

Meanwhile, another kind of high-energy collisions was proposed that requires
the magnetic field \cite{fr}. If the black hole is weakly magnetized but the
charge-to-mass ratio $q/m$ of a particle is big enough, the effect happens
even in the simplest Schwarzschild background, so one can neglect

the backreaction of the magnetic field on the metric. In the absence of a
black hole rotation, there are no special relations between the energy and
the angular momentum, so there is no direct analogue of the critical
particle typical of the Kerr metric. However, kinematic explanation of the
effect considered in \cite{fr} is very similar. One particle moves on the
innermost stable circular orbit (ISCO) that lies close to the horizon for a
sufficiently large magnetic field, the second one being arbitrary, so again
we have separation to two classes of particles.

Quite recently, another version of collisions in the magnetic field, not
connected with ISCO, was proposed in \cite{mpl}. It was shown that there
exists a sharp region near the horizon where collisions in the magnetic
field strong enough lead to unbound $E_{c.m.}$ for \textit{any }two
particles that lends some universality to the effect.

Then, it seems that we have a paradox since we have high $E_{c.m.}$. without
separation to critical and usual particles. Meanwhile, it was found earlier
that collision between two usual particles gives rise to modest $E_{c.m.}$ 
\cite{k}. The aim of the present work is to show that the effect under
discussion in the magnetic field has another underlying kinematic reason as
compared to the original BSW effect. We give below simple kinematic
explanation in the same sense as Ref. \cite{k} gave explanation to the BSW
effect. Throughout the paper we use units in which fundamental constants are 
$G=c=1$.

\section{Metric and equations of motion}

Let us consider the static metric 
\begin{equation}
ds^{2}=-N^{2}dt^{2}+\frac{dr^{2}}{A}+R^{2}d\phi ^{2}+g_{\theta }d\theta ^{2}%
\text{,}
\end{equation}%
where the metric coefficients do not depend on $t$ and $\phi $. The horizon
lies at $r=r_{+}$ and corresponds to $N=0$. We also assume that there is an
electromagnetic field with the four-vector $A^{\mu }$ 
\begin{equation}
A^{\phi }=\frac{B}{2},  \label{ab}
\end{equation}%
all other components $A^{\mu }=0$.

In vacuum, this is an exact solution with $B=const$ \cite{wald}. We consider
configuration with matter (in this sense a black hole is "dirty"), so in
general $B$ may depend on $r$ and $\theta $. We consider motion constrained
within the equatorial plane, so $\theta =\frac{\pi }{2}$. Redefining the
radial coordinate $r\rightarrow \rho $, we can always achieve that 
\begin{equation}
A=N^{2}  \label{an}
\end{equation}%
within this plane. Then, equations of motion read 
\begin{equation}
\dot{t}=\frac{E}{N^{2}m}\text{,}  \label{t}
\end{equation}%
\begin{equation}
\dot{\phi}=\frac{\beta }{R}\text{,}  \label{ft}
\end{equation}%
\begin{equation}
m\dot{r}=\varepsilon Z\text{,}  \label{rt}
\end{equation}%
\begin{equation}
\beta =\frac{L}{mR}-\frac{qBR}{2m}\text{,}  \label{b}
\end{equation}%
\begin{equation}
Z=\sqrt{E^{2}-m^{2}N^{2}(1+\beta ^{2})}\text{,}  \label{v}
\end{equation}

$\varepsilon =-1$ if the particle moves towards the horizon and $\varepsilon
=+1$ in the opposite case. Here, $m$ is the particle's mass, $q$ is its
electric charge.

To describe kinematic properties of particles, it is convenient to introduce
a tangent locally flat space with the use of tetrads. This will enable us to
carry out close analogy with special relativity. Then, the components of the
local three-velocity are equal to%
\begin{equation}
V_{(a)}=V^{(a)}=-\frac{u^{\mu }h_{(a)\mu }}{u^{\mu }h_{(0)\mu }}\text{.}
\label{va}
\end{equation}

Here, $u^{\mu }$ is the four-velocity, $h_{(a)\mu }$ is the tetrad, $\mu
=0,1,2,3$, $(a)=1,2$, $3$. Then, denoting coordinates $x^{\mu }$ as \ $%
x^{0}=t,x^{1}=r$, $x^{2}=\theta $, $x^{3}=\phi $, it is natural to choose
the tetrad in the form $h_{(0)\mu }=-N(1,0,0,0)$, $h_{(1)\mu
}=N^{-1}(0,1,0,0),$ $h_{(2)\mu }=\sqrt{g_{\theta \theta }}%
(0,0,1,0),h_{(3)\mu }=R(0,0,0,1).$ Then, it follows from (\ref{t}) - (\ref{v}%
) that%
\begin{equation}
V^{(1)}=\sqrt{1-\frac{N^{2}}{\alpha ^{2}}(1+\beta ^{2})}  \label{v1}
\end{equation}%
\begin{equation}
V^{(3)}=\frac{\beta N}{\alpha }\text{,}  \label{v3}
\end{equation}%
where%
\begin{equation}
\alpha =\frac{E}{m}\text{.}
\end{equation}%
The component $V^{(2)}=0$ since a particle moves within the equatorial
plane. From (\ref{v1}), (\ref{v3}) it follows that%
\begin{equation}
E=m\gamma _{0}=\frac{mN}{\sqrt{1-V^{2}}}\text{,}  \label{e}
\end{equation}%
where $V^{2}=\left( V^{(1)}\right) ^{2}+\left( V^{(3)}\right) ^{2}$, $\gamma
_{0}$ has the meaning of the Lorentz gamma factor for a given particle.

\section{Particle collisions}

Let two particle collide at some point where $r=r_{c}$. In this point, one
can define the energy in the centre of mass frame according to%
\begin{equation}
E_{c.m.}=-P^{\mu }P_{\mu }=m_{1}^{2}+m_{2}^{2}+2m_{1}m_{2}\gamma \text{,}
\end{equation}%
\begin{equation}
P^{\mu }=m_{1}u_{1}^{\mu }+m_{2}u_{2}^{\mu }\text{,}
\end{equation}%
the Lorentz factor of relative motion (not to be confused with the factor $%
\gamma _{0}$ for each individual particle)%
\begin{equation}
\gamma =-u_{1\mu }u_{2}^{\mu }\text{,}  \label{g}
\end{equation}%
indices 1 and 2 label the particles. Then, using (\ref{t}) - (\ref{v}), one
finds%
\begin{equation}
\gamma =\frac{E_{1}E_{2}-\varepsilon _{1}\varepsilon _{2}Z_{1}Z_{2}}{%
N^{2}m_{1}m_{2}}-\beta _{1}\beta _{2}\text{.}  \label{ga}
\end{equation}

Let us define the function $s(r)=N\beta $, $s_{i}$ $s_{i}\equiv s_{i}(r_{c})$
for particle $i$. We assume that at least for one of two particles%
\begin{equation}
s_{i}\sim 1.\text{ }  \label{s}
\end{equation}%
This means that the large $\beta $ compensates the small factor $N$ in the
point of collision near the horizon. Then, the gamma factor reads \cite{mpl}%
\begin{equation}
\gamma \approx \frac{F}{N^{2}(r_{c})}\text{, }F=\alpha _{1}\alpha _{2}-\sqrt{%
\alpha _{1}^{2}-s_{1}^{2}}\sqrt{\alpha _{2}^{2}-s_{2}^{2}}-s_{1}s_{2},
\label{gas}
\end{equation}%
it grows unbound near the horizon.

If near the horizon we neglect a small difference between $\beta (r)$ and $%
\beta (r_{+})$, 
\begin{equation}
s_{i}(r)\approx s_{i}\frac{N(r)}{N(r_{c})}\text{.}
\end{equation}%
Thus $s_{i}(r_{+})=0$. The unbound $\gamma $ is possible in the strip near
the horizon where $s_{i}\leq \alpha _{i}$ or, equivalently,%
\begin{equation}
0<N_{c}\leq \frac{E_{i}}{m_{i}\beta (r_{+})}\text{.}  \label{str}
\end{equation}

In other words, for any value of magnetic field large enough, so that $%
\left\vert \beta (r_{+})\right\vert \gg 1$, there exist a narrow strip
around horizon within which $\gamma \sim N_{c}^{-2}\sim B^{2}$ is large.

The mechanism under discussion works both for the nonextremal ($N^{2}\sim
(r-r_{+}))$ and extremal ($N^{2}\sim (r-r_{+})^{2}$) cases. In other words,
we require $N$ to be small in the point of collision near the horizon but do
not specify the rate with which it vanishes.

It is worth noting that for collisions near ISCO, self-gravitation of very
strong magnetic fields ($BM\gtrsim 1$, where $M$ is a black hole mass)
restricts the growth of $E$ $_{c.m.}$ (see Sec. XII of \cite{in}) since the
location of ISCO fails to be close to the horizon. By contrast, in the case
under discussion the effect is not tight to ISCO, so high energies are
achievable as long as eq. (\ref{s}) is satisfied in the strong magnetic
field near the horizon .

These features give more universality to the effect under discussion.

\section{Properties of velocities}

Properties of the gamma factor can be understood in terms of the relative
velocity $w$ of two particles:%
\begin{equation}
\gamma =\frac{1}{\sqrt{1-w^{2}}}\text{,}  \label{gw}
\end{equation}%
The unbound growth of $\gamma $ (hence, the BSW effect) happens when $%
w\rightarrow 1$. If in the flat space-time particles have the
three-velocities $\vec{V}_{1}=V_{1}\vec{n}_{1}$ and $\vec{V}_{2}=V_{2}\vec{n}%
_{2}$,%
\begin{equation}
w^{2}=1-\frac{(1-V_{1}^{2})(1-V_{2}^{2})}{[1-V_{1}V_{2}(\vec{n}_{1}\vec{n}%
_{2})]^{2}},  \label{w}
\end{equation}%
see for details problem 1.3. in \cite{lt}. Possible cases in which $w$
approaches $1$ (speed of light in our units) can be classified depending on $%
V_{1}$, $V_{2}$ and the angle between them, determined by $(\vec{n}_{1}\vec{n%
}_{2})$. For the standard BSW effect, the only relevant case occurs when $%
V_{1}<1$, $V_{2}\rightarrow 1$, $(\vec{n}_{1}\vec{n}_{2})$ is arbitrary
(case "a" according to classification given in Sec. III of Ref. \cite{k}).
Then, it is clear from (\ref{w}) that indeed $w\rightarrow 1$.

However, now eq. (\ref{e}) tells us that for any finite $E$, the velocity $%
V\rightarrow 1$ on the horizon, so the critical particle in the sense of 
\cite{k} is impossible. It was possible for rotating black holes since for
them the left hand side of eq. (\ref{e}) is to be replaced with $E-\omega L$%
, where the metric coefficient $\omega $ is responsible for rotation, so for
the critical particle $E=\omega _{H}L$, where $\omega _{H}$ is the angular
velocity of a black hole. But now $\omega _{H}=0$, $E>0$. Happily, there is
one more case that ensures the desirable limit of $w$. Let $V_{1}\rightarrow
1$, $V_{2}\rightarrow 1$ in such a way that 
\begin{equation}
V_{i}=1-A_{i}\delta  \label{vi}
\end{equation}%
where $A_{i}$ ($i=1,2)$ are constants, $\delta \ll 1$, 
\begin{equation}
(\vec{n}_{1}\vec{n}_{2})\neq 1.  \label{n}
\end{equation}%
This is case b1 from \cite{k}. Then, it follows from (\ref{w}) that%
\begin{equation}
w^{2}\approx 1-\frac{4A_{1}A_{2}\delta ^{2}}{[1-(\vec{n}_{1}\vec{n}_{2})]^{2}%
}\text{,}  \label{w1}
\end{equation}%
so we have $w\rightarrow 1$ again.

Now we will see that for collisions in the magnetic field under discussion
it is this case which is realized. To demonstrate this, we must check two
conditions - (\ref{vi}) and (\ref{n}). The validity of (\ref{vi}) follows
immediately from (\ref{e}): for both particles $V_{1}=1-O(N^{2})$, so they
approach 1 with the same rate as required.

Now, we must check (\ref{n}). It is instructive to compare two cases - (i)
the magnetic field is absent or at least finite, so $\beta (r_{+})\lesssim 1$
(ii) condition (\ref{s}) is satisfied, so $\beta (r_{+})\gg 1$.

In case (i) typical of the standard BSW effect \cite{ban}, \cite{k}, in the
horizon limit $N\rightarrow 0$ we obtain that $V^{(1)}\rightarrow 1$, $%
V^{(3)}\rightarrow 0$. It means that when a particle crosses the horizon,
its velocity is pointed always in the same direction. Namely, it is
perpendicular to the horizon for any particle, so that $(\vec{n}_{1}\vec{n}%
_{2})\rightarrow 1$: rains falls vertically on a black hole - see p. F-17 of 
\cite{tw}. Therefore, $(\vec{n}_{1}\vec{n}_{2})=1$, so (\ref{n}) is violated.

In case (ii), it follows from (\ref{s}), (\ref{v1}), (\ref{v3}) that%
\begin{equation}
V^{(1)}(r_{c})\approx \varepsilon \sqrt{1-\frac{s^{2}}{\alpha ^{2}}}\text{,}
\label{1s}
\end{equation}%
\begin{equation}
V^{(3)}(r_{c})\approx \frac{s}{\alpha }\text{.}  \label{3s}
\end{equation}%
Thus the angle with which they approach the horizon can be arbitrary and
different for different particles, so that (\ref{n}) is satisfied. More
precisely, one can find from (\ref{v1}), (\ref{v3}) that for $\varepsilon
_{1}=\varepsilon _{2}=-1$,%
\begin{equation}
(\vec{n}_{1}\vec{n}_{2})=\frac{\sqrt{\alpha _{1}^{2}-s_{1}^{2}}\sqrt{\alpha
_{2}^{2}-s_{2}^{2}}+s_{1}s_{2}}{\alpha _{1}\alpha _{2}}\text{.}
\end{equation}%
From (\ref{e}), we have%
\begin{equation}
1-V_{i}^{2}=\frac{N^{2}}{\alpha _{i}^{2}}\text{.}
\end{equation}%
Then, using (\ref{gw}) and (\ref{w}), one finds just the expression (\ref%
{gas}) that was derived in \cite{mpl} from the equations of motion.

It is worth noting that this is valid for the point of collision, where the
product $N\beta $ is finite according to (\ref{s}). However, for any large
but finite magnetic field (so $\beta $ is finite as well), this product
vanishes on the horizon itself, so again $(\vec{n}_{1}\vec{n}%
_{2})\rightarrow 1$, $V\rightarrow 1$ for both particles. Although,
according to (\ref{str}), in the close vicinity of the, say, nonextremal
horizon $r_{c}-r\sim b^{-2}$ , "rain does not fall vertically", its
direction is restored when $r-r_{+}\ll r_{c}-r$, \ so $s(r)\ll 1$.

In accordance with the general rule that collision of two usual particles
does not produce unbound $E_{c.m.}$\cite{k}, the Lorentz factor $\gamma $
becomes finite. Indeed, it follows from (\ref{ga}) that $\lim_{r\rightarrow
r_{+}}\gamma \sim \beta ^{2}(r_{+}).$ However, if the magnetic field is
strong enough, $\gamma $ can be made as large as one likes.

\section{Summary}

Thus the original BSW effect and high-energy collision in the magnetic field
in the immediate vicinity of the horizon \cite{mpl} have two complimentary
kinematic explanations. They correspond to cases (a) and (b) of eqs. 19 - 20
in \cite{k}. Namely, in case (a) high $E_{c.m.}$ are achieved when one
particle is critical (fine-tuned, slow) and the second particle is usual
(not fine-tuned, rapid). For scenarios of the type considered in \cite{mpl},
both particles are usual and rapid, big $E_{c.m.}$ are obtained since their
velocities are non-parallel to each other. Thus the kinematic explanations
of both processes (without the magnetic field and in the strong magnetic
field) prove to be two different versions of the same underlying picture.

\section{Acknowledgement}

This work was founded by the subsidy allocated to Kazan Federal University
for the state assignment in the sphere of scientific activities.


\begin{thebibliography}{9}
\bibitem{ban} M. Ba\~{n}ados, J. Silk and S.M. West, Phys. Rev. Lett. 
\textbf{103} (2009) 111102 [arXiv:0909.0169].98].

\bibitem{k} O. B. Zaslavskii, Phys. Rev. D \textbf{84}, 024007 (2011)
[arXiv:1104.4802].

\bibitem{fr} V. P. Frolov, Phys. Rev. D\textbf{\ 85}, 024020 (2012)
[arXiv:1110.6274].

\bibitem{mpl} O. B. Zaslavskii, Mod. Phys. Lett. A \textbf{29}, 1450112
(2014) [arXiv:1403.6286].

\bibitem{wald} R. M. Wald, Phys. Rev. D \textbf{10}, 1680 (1974).

\bibitem{in} O. B. Zaslavskii, [arXiv:1405.2543].

\bibitem{lt} A. P. Lightman, W. H. Press, R. H. Price, and S. A. Teukolsky,%
\textit{\ Problem book in Relativity and Gravitation }(Princeton University
Press, Princeton, New Jersey, 1975).

\bibitem{tw} E. F. Taylor and J. A. Wheeler, \textit{Exploring Black Holes:
Introduciton to general relativity} (Addison Wesley Longman, Inc., 2000).
\end{thebibliography}
\end{document}